\documentclass[%
 reprint,
 superscriptaddress,
 showpacs,
 amsmath,amssymb,
 aps,
 prl,
]{revtex4-1}

\usepackage{graphicx}
\usepackage{dcolumn}

\usepackage{siunitx}
\usepackage{bm}
\usepackage{upgreek}

\usepackage[symbol*]{footmisc}
\DefineFNsymbolsTM{otherfnsymbols}{%
	\textdagger            \dagger
	\textasteriskcentered  *
}%
\setfnsymbol{otherfnsymbols}

\begin{document}

\preprint{APS/123-QED}

\title{Polarization constraints in reciprocal unitary backscattering} 

\affiliation{School of Electrical and Electronic Engineering, Nanyang Technological University, Singapore, 639798, Singapore.}
\affiliation{School of Chemical and Biomedical Engineering, Nanyang Technological University, Singapore, 637459, Singapore.}
\affiliation{Harvard Medical School and Massachusetts General Hospital, Wellman Center for Photomedicine, Boston, Massachusetts 02114, USA.}
\affiliation{Institute for Medical Engineering and Science, Massachusetts Institute of Technology, Cambridge, Massachusetts 02139, USA}

\author{Qiaozhou \surname{Xiong}}
	\thanks{These authors contributed equally}
	\affiliation{School of Electrical and Electronic Engineering, Nanyang Technological University, Singapore, 639798, Singapore.}
\author{Nanshuo \surname{Wang}} %
	\thanks{These authors contributed equally}
	\affiliation{School of Electrical and Electronic Engineering, Nanyang Technological University, Singapore, 639798, Singapore.}
\author{Xinyu \surname{Liu}}
	\affiliation{School of Electrical and Electronic Engineering, Nanyang Technological University, Singapore, 639798, Singapore.}
\author{Si \surname{Chen}}
	\affiliation{School of Electrical and Electronic Engineering, Nanyang Technological University, Singapore, 639798, Singapore.}
\author{Cilwyn S. \surname{Braganza}}
	\affiliation{School of Electrical and Electronic Engineering, Nanyang Technological University, Singapore, 639798, Singapore.}
\author{Brett E. \surname{Bouma}}
	\affiliation{Harvard Medical School and Massachusetts General Hospital, Wellman Center for Photomedicine, Boston, Massachusetts 02114, USA.}
	\affiliation{Institute for Medical Engineering and Science, Massachusetts Institute of Technology, Cambridge, Massachusetts 02139, USA}

\author{Linbo \surname{Liu}}
	\thanks{These authors contributed equally.\\\begin{tabular}{l l} Corresponding authors: & LIULINBO@ntu.edu.sg\\{ } & MVILLIGER@mgh.harvard.edu\end{tabular}}
	\affiliation{School of Electrical and Electronic Engineering, Nanyang Technological University, Singapore, 639798, Singapore.}
	\affiliation{School of Chemical and Biomedical Engineering, Nanyang Technological University, Singapore, 637459, Singapore.}
	
\author{Martin \surname{Villiger}}
	\thanks{These authors contributed equally.\\\begin{tabular}{l l} Corresponding authors: & LIULINBO@ntu.edu.sg\\{ } & MVILLIGER@mgh.harvard.edu\end{tabular}}
	\affiliation{Harvard Medical School and Massachusetts General Hospital, Wellman Center for Photomedicine, Boston, Massachusetts 02114, USA.}

\date{\today}

\begin{abstract}

We observed that the polarization state of light after round-trip propagation through a birefringent medium frequently aligns with the employed input polarization state ``mirrored'' by the horizontal plane of the Poincar\'{e} sphere. In this letter we explore the predisposition for this mirror state and demonstrate how it constrains the evolution of polarization states as a function of the round-trip depth into weakly scattering birefringent samples, as measured with polarization-sensitive optical coherence tomography (PS-OCT). The constraint enables measurements of depth-resolved sample birefringence with PS-OCT using only a single input polarization state, which offers a critical simplification compared to the use of multiple input states.


\end{abstract}

\maketitle



Polarization offers access to unique, distinguishing signatures of samples for diverse applications from remote sensing [\onlinecite{sassen1991polarization,tyo2006review}] to biomedical optics [\onlinecite{ghosh2011tissue,tuchin2016polarized,qi2017mueller}]. Conventionally, multiple input polarization states are required in addition to polarization-diverse detection to fully characterize the polarization properties of a sample, prompting complex instrumentation. Alleviating these hardware requirements would enable more widespread exploration of this compelling contrast mechanism. 

\begin{figure}
\includegraphics[scale=1]{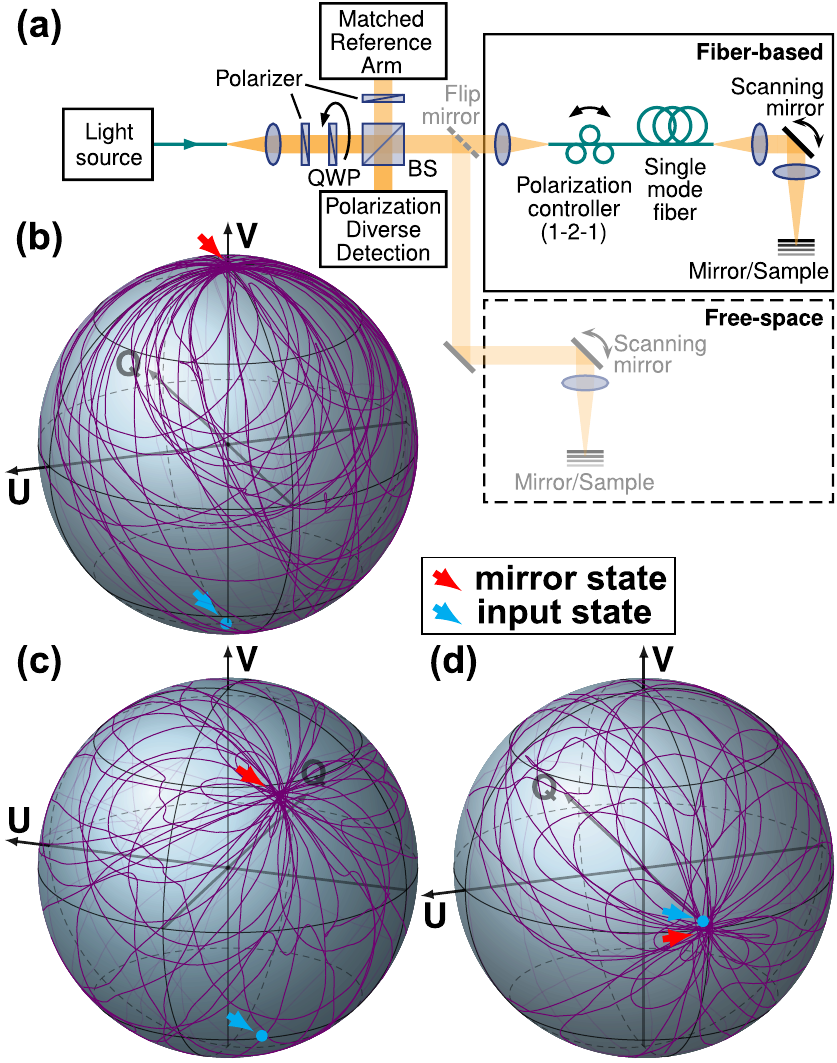}
\caption{\label{fig:fig1} Demonstration of the polarization mirror state. (a) Schematic drawing of the optical system employed throughout all experiments. QWP: quarter-wave plate; BS: beamsplitter. (b)-(d) Polarization state evolution on the Poincar\'{e} sphere as a result of moving the three paddles of the polarization controller when using circularly, elliptically and linearly polarized input states, respectively, indicated by the blue arrows.}
\end{figure}

In previous experiments, we observed that the polarization state of backscattered or reflected light, when measured through identical illumination and detection paths, frequently evolved through the employed input polarization state but with reversed handedness, corresponding to the input state mirrored by the horizontal plane of the Poincar\'{e} sphere [\onlinecite{wang2017polarization}]. Earlier investigations of the polarization properties of single mode fibers reported on aspects of the polarization mirror state [\onlinecite{brinkmeyer1981forward,van1993polarization,corsi1998polarization}], yet without elucidating its manifestation. To examine the polarization mirror state, we measured the round-trip signal through a 1.5-\si{\meter}-long single-mode optical fiber. Instead of using a conventional polarimeter, we employed interferometric measurements for the later coherence gating experiments, as depicted in Fig.~\ref{fig:fig1}(a). Light from a super-continuum source was linearly polarized, prepared with an achromatic quarter-wave plate (QWP) to different input polarization states, and split into reference and sample arms. A linear polarizer oriented at \ang{45} in the reference arm defined the reference polarization state independent of the input polarization. For polarization diverse detection, the sample and reference light was combined in a polarization-maintaining fiber to then direct each of the fiber’s two linear eigenstates towards a grating-based spectrometer (760--\SI{920}{nm}). The recorded fringe signals reveal the amplitude and relative phase of the two orthogonal electromagnetic field components in the sample arm and hence the polarization state of the sample light.
 
Employing a polarization controller to alter the birefringence of the fiber, we measured the time-varying polarization state resulting from randomly moving the paddle positions of the controller. Visualized in Fig.~\ref{fig:fig1}(b) as the normalized Stokes vectors of the center wavelength (\SI{840}{nm}) in the $Q$, $U$, and $V$-coordinates of the Poincar\'{e} sphere, the polarization mirror state manifests by the repeated crossing of the polarization state evolution in a specific state ${\bf{m}}$, highlighted by the red arrow in Fig.~\ref{fig:fig1}(b). Repeated with different launching polarization states ${\bf{s}}$ (indicated by the blue arrows in Figs.~\ref{fig:fig1}(c,d)), we recognize that ${\bf{m}}={\bf{D}}\cdot{\bf{s}}$, where ${\bf{D}}=\operatorname{diag}(1,1,-1)$. ${\bf{m}}$ corresponds to the input state mirrored by the horizontal $QU$-plane, explaining its designation as the polarization mirror state. The input states were determined by reflecting the light to the detector in free space, without the fiber in place. All measurements were performed in the fixed coordinates of the receiver and are independent of the orientation of the coordinates in the illumination path.

\begin{figure}
\includegraphics[scale=1]{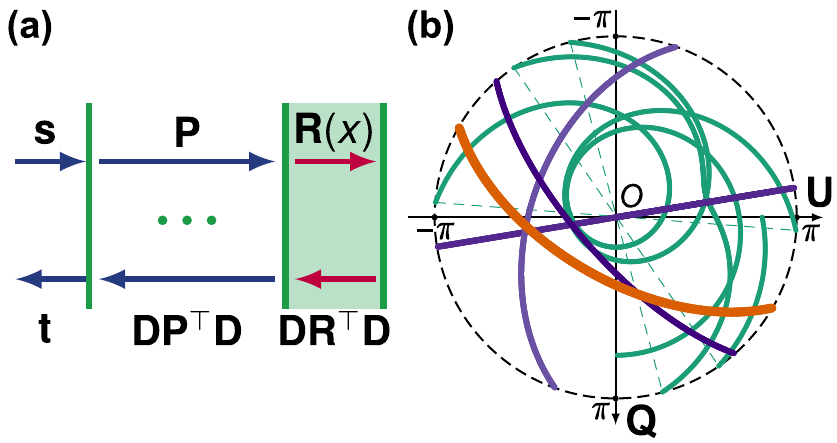}
\caption{\label{fig:fig2} Theoretical explanation of the polarization mirror state. (a) Model of the round-trip propagation through a reciprocal sample, comprising ${\mathbf{R}}(x)$ with varying retardation and a static element ${\mathbf{P}}$. (b) The rotation vectors of all linear retarders localize within a circle of radius $\pi$ within the $QU$-plane of the Poincar\'{e} sphere. The orange curve represents the end points of the rotation vectors mapping the randomly chosen input polarization state $\left[ \cos(-\pi/6)\cdot\cos(\pi/3), \cos(-\pi/6)\cdot\sin(\pi/3), \sin(-\pi/6)\right]^\top$ exactly to its mirror point. The green trace represents the simulated rotation vector evolution for a synchronous movement of the polarization controller paddles. The purple curves represent the rotation vectors of ${\mathbf{D}}\cdot{\mathbf{P}}^\top\cdot{\mathbf{R}}^\top(x)\cdot{\mathbf{D}}\cdot{\mathbf{R}}(x)\cdot{\mathbf{P}}$, where ${\mathbf{R}}(x)$ is a linearly increasing linear retarder, for three representative sets of distinct ${\mathbf{P}}$ and ${\mathbf{R}}(x)$.}
\end{figure}

To appreciate the mirror state phenomenon, we consider a general retarder ${\mathbf{R}}(x)$ with its retardation varying as a function of $x$, e.g. the polarization controller's paddle positions. The retarder may be preceded by a static element ${\mathbf{P}}$. The combined system, illustrated in Fig.~\ref{fig:fig2}(a), transforms the input polarization state ${\mathbf{s}}$ into the output state ${\mathbf{t}}$:

\begin{equation}
{\mathbf{t}} = {\mathbf{D}} \cdot {\mathbf{P}}^\top \cdot {\mathbf{R}}^\top (x) \cdot {\mathbf{D}} \cdot 
{\mathbf{R}}(x) \cdot {\mathbf{P}} \cdot {\mathbf{s}} = {\mathbf{T}}(x) \cdot {\mathbf{s}}
\label{eqn:eq1}
\end{equation}

Here, ${\mathbf{T}}={\mathbf{D}} \cdot {\mathbf{P}}^\top \cdot {\mathbf{R}}^\top \cdot {\mathbf{D}} \cdot {\mathbf{R}} \cdot {\mathbf{P}}$, where $^\top$ denotes transpose, and all vectors and matrices are in the rotation group SO(3). We chose to follow the convention of maintaining the orientation of the spatial $xy$-coordinates irrespective of the light’s propagation direction [\onlinecite{vansteenkiste1993optical,cloude1995concept}]. In reciprocal media, the reverse transmission through element ${\mathbf{R}}$ is described by ${\mathbf{D}}\cdot{\mathbf{R}}^\top\cdot{\mathbf{D}}$ [\onlinecite{potton2004reciprocity,gil2007polarimetric}] (see supplementary material section 1 [\onlinecite{supplementary}]). It is important to note that the round-trip transmission ${\mathbf{T}}$ is $D$-transpose symmetric ${\mathbf{T}}={\mathbf{D}}\cdot{\mathbf{T}}^\top\cdot{\mathbf{D}}$, which makes ${\mathbf{T}}$ a linear retarder. The round-trip effectively cancels any optical activity or circular retardation and relates to the weak localization of light [\onlinecite{van1988polarisation}]. The effect of ${\mathbf{T}}$ on the input state can be described by a rotation vector ${\boldsymbol{\uptau}}(x)$ lying in the $QU$-plane of the Poincar\'{e} sphere, with its direction indicating the rotation axis, and its length defining the amount of rotation. Considering their $2\pi$-ambiguity, the rotation vectors of all possible linear retarders are confined to a circle with a radius of $\pi$ within the $QU$-plane (Fig.~\ref{fig:fig2}(b)). When moving the polarization controller paddles, ${\boldsymbol{\uptau}}(x)$ traces out an intricate path in the $QU$-plane, as shown by the green line in Fig.~\ref{fig:fig2}(b) for simulating a synchronous movement of the three paddles. 

There exists only a single rotation vector within the $QU$-plane that rotates a given input state ${\mathbf{s}}$ onto an arbitrary output state ${\mathbf{t}}$. This rotation vector is defined by the intersection of the $QU$-plane and the plane bisecting ${\mathbf{s}}$ and ${\mathbf{t}}$. In order for ${\mathbf{s}}$ to pass through ${\mathbf{t}}$, ${\boldsymbol{\uptau}}(x)$ has to evolve through this specific point within the $\pi$-circle of the $QU$-plane. The only exception, there exists a continuum of rotation vectors that map ${\mathbf{s}}$ onto its mirror state ${\mathbf{m}}={\mathbf{D}}\cdot{\mathbf{s}}$, because the $QU$-plane coincides with the bisecting plane in this case. These rotation vectors are located on a curve ${\boldsymbol{\uptau}}_{\mathbf{m}}$ within the $QU$-plane (orange curve in Fig.~\ref{fig:fig2}(b), and supplementary material section 2 [\onlinecite{supplementary}]). Every intersection of ${\boldsymbol{\uptau}}(x)$ with ${\boldsymbol{\uptau}}_{\mathbf{m}}$ corresponds to ${\mathbf{s}}$ evolving through the mirror state ${\mathbf{m}}$, explaining its frequent realization.

Importantly, the presence of diattenuation that induces polarization-dependent loss would skew the measured polarization states and frustrate the repeated evolution through the mirror state. The mirror state only manifests in systems that can be accurately modeled with unitary transmission matrices.

We next used PS-OCT to measure the polarization state of light backscattered within a scattering sample as a function of its round-trip depth [\onlinecite{hee1992polarization,de2017polarization}]. At the scale of the axial resolution of OCT, tissue can be modeled as a sequence of homogeneous linearly birefringent layers with distinct optic axis orientations. ${\mathbf{R}}(x)$ describes in this case a linear retarder with a retardance that linearly increases with depth $x$, resulting in ${\mathbf{D}}\cdot{\mathbf{R}}^\top\cdot{\mathbf{D}}={\mathbf{R}}$. ${\mathbf{P}}$ contains the combined effect of system components and preceding tissue layers. The resulting rotation vectors ${\boldsymbol{\uptau}}(x)$ form regular curves across the $\pi$-circle (purple curves in Fig.~\ref{fig:fig2}(b)). All possible traces intersect the curve ${\boldsymbol{\uptau}}_{\mathbf{m}}$ precisely once, ensuring periodic crossing of ${\mathbf{m}}$. To inspect in more detail the evolution of ${\mathbf{t}}$, we take its derivative with respect to $x$, and substitute ${\mathbf{s}}={\mathbf{T}}^\top(x)\cdot{\mathbf{t}}$:

\begin{equation}
\frac { \partial {\mathbf{t}} } { \partial x } = \frac { \partial {\mathbf{T}} ( x ) } { \partial x } \cdot {\mathbf{T}}^\top (x) \cdot {\mathbf{t}} = {\boldsymbol{\upbeta}} (x) \times {\mathbf{t}}
\label{eqn:eq2}
\end{equation}

Because ${\mathbf{T}}^\top\cdot{\mathbf{T}}$ is the identity matrix, $\left( \partial{\mathbf{T}} / \partial x \right) \cdot {\mathbf{T}}^{\top} = - {\mathbf{T}} \cdot \left( \partial {\mathbf{T}}^{\top} / \partial x \right)$ is skew-symmetric and can be expressed as the cross-product operator ${\boldsymbol{\upbeta}} \times$, which is constant for a retardance that linearly increases with $x$ (see supplementary material section 3 [\onlinecite{supplementary}]). Accordingly, within a single sample layer, ${\mathbf{t}}$ evolves on the Poincar\'{e} sphere with constant speed rotating around the apparent optic axis ${\boldsymbol{\upbeta}}$ on a circle constrained to pass through ${\mathbf{m}}$.

\begin{figure}
\includegraphics[scale=1]{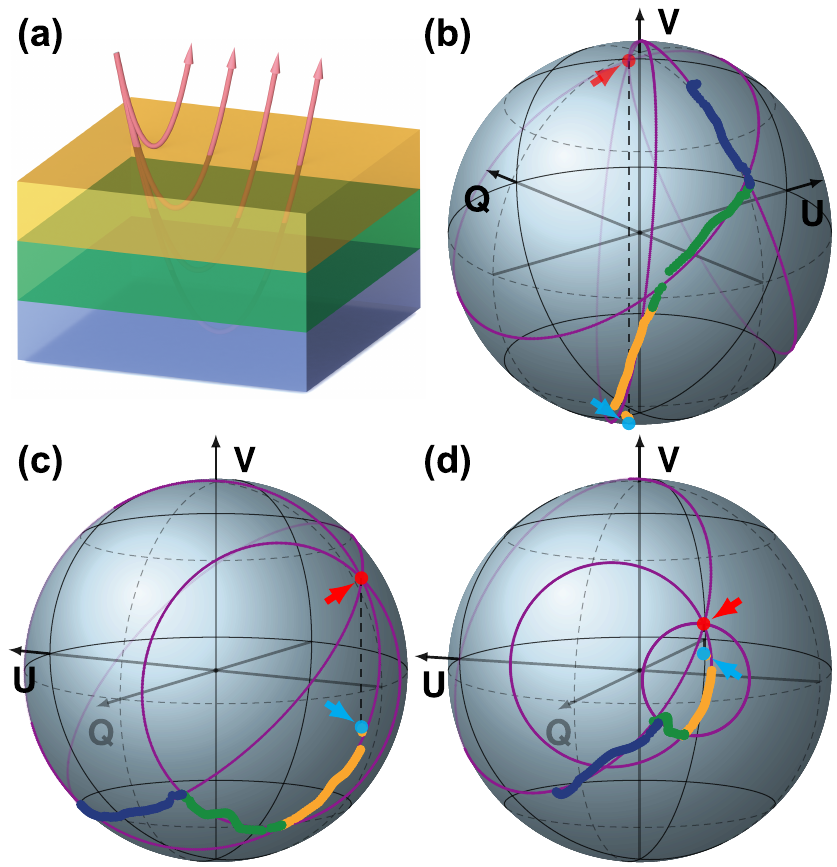}
\caption{\label{fig:fig3} Evolution of coherence-gated polarization states in a three-layer birefringence phantom. (a) Schematic sketch of the phantom consisting of three layers with distinct optic axis orientations. (b)-(d) Polarization state evolution (color-coded corresponding to the layers in (a)) for a circularly polarized (b), elliptically polarized (c), and linearly polarized input state (d).}
\end{figure}

For experimental validation, we prepared a scattering phantom consisting of three linearly birefringent layers with distinct optic axis orientations [\onlinecite{liu2017tissue}] (Fig.~\ref{fig:fig3}(a)). Without the fiber segment in the sample arm, we focused the light with a \SI{30}{mm} focal length lens into the sample, achieving a full-width at half maximum (FWHM) spot diameter of \SI{\sim 8}{\mu m}, and scanned with galvanometric mirrors in the lateral direction. The spectrometer's bandwidth offers an axial resolution of \SI{\sim 2.4}{\mu m}. At each scanning location, using PS-OCT, we constructed the Stokes vector as a function of depth in the sample. To remove speckle and improve the signal, we spatially filtered the original Stokes vectors with a two-dimensional Gaussian kernel of \SI{20}{\mu m} $1/{e^2}$ width in the axial direction and \SI{80}{\mu m} in the lateral direction. Finally, we computed the normalized three-component Stokes vector ${\mathbf{r}}(z)$ as a function of depth, shown in Figs.~\ref{fig:fig3}(b-d) for three distinct input polarization states at one lateral sample location. We then fitted circles to the polarization state evolution within each layer. The circles (in purple color) demonstrate a close match with the measured polarization states and all circles evolve through the polarization mirror state ${\mathbf{m}}$ (indicated by red arrows), as expected.

Using a single input polarization state for PS-OCT, it is straightforward to compute the cumulative retardation that propagation through the sample to a given depth and back imparts on the input polarization state [\onlinecite{hitzenberger2001measurement}]. Yet, cumulative retardation can be difficult to interpret in samples with a layered architecture, and it is more insightful to compute local retardation, i.e. the derivative of the retardance of ${\mathbf{T}}(x)$ with depth, which is given by the norm of $ {\boldsymbol{\upbeta}} $ and is proportional to the sample birefringence [\onlinecite{guo2004depth,todorovic2004determination,makita2010generalized}] at that depth location. Following Eq.~(\ref{eqn:eq2}) we have

\begin{equation}
\left| \frac { \partial {\mathbf{t}} } { \partial x } \right| = \left| {\boldsymbol{\upbeta}} \right| \sin \theta
\label{eqn:eq3}
\end{equation}

where we used $\left|  {\mathbf{t}} \right|=1$. $\theta$ is the angle between the rotation vector ${\boldsymbol{\upbeta}}$ and the polarization state ${\mathbf{t}}$ and is needed to deduce local retardation. With only a single input state this angle is generally unknown. Using, instead, two input polarization states oriented at \SI{90}{\degree} to each other on the Poincar\'{e} sphere reveals the orientation of the apparent optic axis. However, recognizing that the evolution of ${\mathbf{t}}$ is constrained to go through ${\mathbf{m}}$, it is possible to recover the orientation and magnitude of ${\boldsymbol{\upbeta}}$ from measurements with only a single input polarization state. Owing to this constraint, both $\partial {\mathbf{t}} / \partial x$ and $\left( {\mathbf{t}}-{\mathbf{m}} \right)$ lie within the same plane orthogonal to ${\boldsymbol{\upbeta}}$. Hence, the direction of ${\boldsymbol{\upbeta}}$ can be obtained by the cross-product ${\boldsymbol{\upbeta}}_0={\left( \partial {\mathbf{t}} / \partial x \right)} \times {\left( {\mathbf{m}}-{\mathbf{t}} \right)}$, and $\sin \theta={\left|{\boldsymbol{\upbeta}}_0 \times {\mathbf{t}} \right|}/{\left| {\boldsymbol{\upbeta}}_0 \right|}$, allowing to calculate, after some algebraic manipulations:

\begin{equation}
{\boldsymbol{\upbeta}}=\dfrac{\dfrac{\partial {\mathbf{t}}}{\partial x} \times {\left( {\mathbf{m}}-{\mathbf{t}} \right)}}{1-{\mathbf{t}}^\top\cdot{\mathbf{m}}}
\label{eqn:eq4}
\end{equation}


\begin{figure*}
\includegraphics[scale=1]{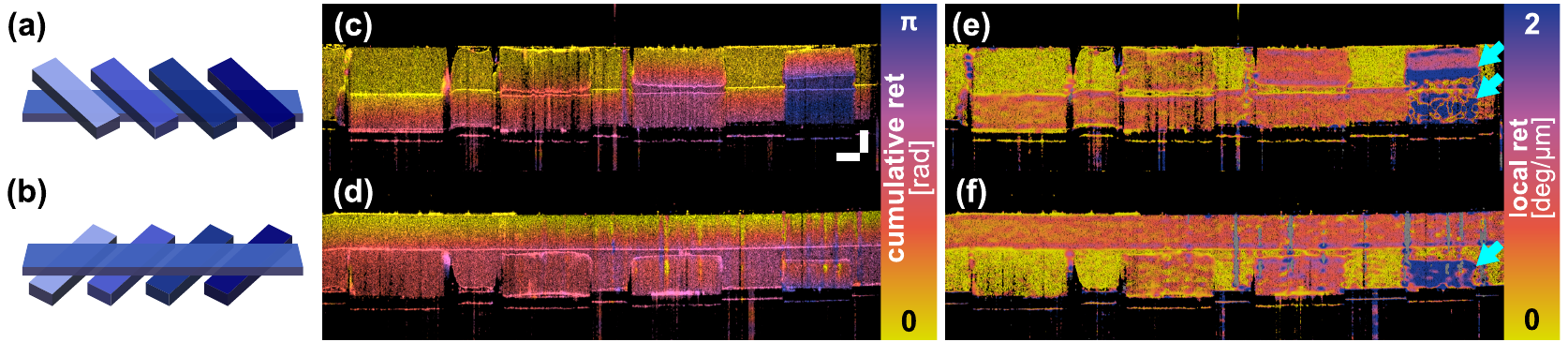}
\caption{\label{fig:fig4} Local retardation imaging of birefringence phantom using the polarization mirror state constraint. (a,b) Schematic drawings of the two-layer phantom in either orientation. (c,d) Corresponding cumulative retardation and (e,f) local retardation images, respectively. Scale bars in (c) measure \SI{100}{\mu m} (vertical) and \SI{400}{\mu m} (horizontal).}
\end{figure*}

To validate the ability of the polarization mirror state to reconstruct local retardation, we imaged a tissue-like phantom consisting of a long birefringent band followed by four parallel elements with distinct birefringence levels and an optic axis orientation different from the long band [\onlinecite{liu2017tissue}] (Fig.~\ref{fig:fig4}(a) and \ref{fig:fig4}(b)) and the gap was filled with non-birefringent matrix.

For reconstruction of local retardation, we employed the pre-calibrated polarization mirror state ${\mathbf{m}}$, and implemented Eq.~(\ref{eqn:eq4}) by approximating ${\mathbf{t}} = \left({\mathbf{r}}\left[p+1\right] + {\mathbf{r}}\left[p\right]\right)/2$ and ${\partial{\mathbf{t}} / \partial x}={\left({\mathbf{r}}\left[p+1\right] - {\mathbf{r}}\left[p\right]\right)}/{\Delta z}$, where $p$ is the pixel index along depth $z = p\cdot\Delta z$, and $\Delta z$ is the axial sampling distance. To avoid high-frequency noise introduced by taking the difference between adjacent points, we axially averaged the reconstructed rotation vector ${\boldsymbol{\upbeta}}(z)$ with a Gaussian window of the same axial size as used to filter the Stokes vectors. The norm of ${\boldsymbol{\upbeta}}$, scaled to degrees of retardation per depth (\si{\degree / \mu m}), reveals the sample's local retardation, imaged with either side of the sample facing up (Figs.~\ref{fig:fig4}(e) and \ref{fig:fig4}(f)). For comparison, the cumulative retardation of ${\mathbf{T}}(x)$ was computed by evaluating the angle between ${\mathbf{r}}(z)$ at each depth and ${\mathbf{r}}(z_{\text{surf}})$, where $z_{\text{surf}}$ is the axial location of the sample surface within each depth profile (Figs.~\ref{fig:fig4}(c) and \ref{fig:fig4}(d)). Whereas cumulative retardation is difficult to interpret, the local retardation clearly reveals the individual sample segments with their distinct levels of birefringence and is recovered irrespective of the sample orientation [\onlinecite{makita2010generalized,villiger2016deep}]. To demonstrate local retardation imaging in biological tissue, we measured {\emph{ex vivo}} swine retina (Supplementary material section 4 [\onlinecite{supplementary}]). 

Previous strategies to reconstruct local birefringence from single-input-state PS-OCT rely on the intrinsic symmetry of the imaging system [\onlinecite{todorovic2004determination,fan2012mapping}] and assume that the optical elements in the illumination and detection paths have no impact on the polarization states. Most OCT instruments for clinical applications, however, use fiber-based optical components with distinct illumination and detection paths, which breaks the intrinsic $D$-transpose symmetry [\onlinecite{villiger2018optic}]. Crucially, the evolution of ${\mathbf{t}}$ through the mirror state persists also in systems with distinct retardation in the illumination and detection optics. This is equivalent to left-multiplying Eq.~(\ref{eqn:eq1}) with an additional matrix ${\mathbf{B}}$. Although the apparent cumulative retarder that maps the input state onto the measured output state is no longer a linear retarder in this case, ${\mathbf{B}}$ simply alters the location of the circular evolution of the polarization states on the Poincar\'{e} sphere to go through the actual mirror state to ${\mathbf{B}}\cdot{\mathbf{m}}$.


A remaining challenge manifests whenever ${\mathbf{t}}$ aligns with ${\mathbf{m}}$, which impairs the reconstruction of local retardation (cyan arrows in Fig.~\ref{fig:fig4}(f)). This corresponds to the effective polarization state in the target layer to orient along one of that layer’s optic axes, and even prevents the cumulative retardation from accumulating retardance. Using circularly polarized input light requires a half wave of retardation to realize this alignment, which is uncommon in many biological samples. Yet, some tissues feature substantial birefringence and controlling the input state is not necessarily possible. The resulting artifact can be avoided by introducing a modest amount of polarization mode dispersion (PMD) into the system and using spectral binning for reconstruction [\onlinecite{villiger2013spectral}]. Because PMD disperses the input polarization state across the spectral bins, simultaneous alignment of ${\mathbf{t}}$ with ${\mathbf{m}}$ in all bins is very unlikely.

\begin{figure}
\includegraphics[scale=1]{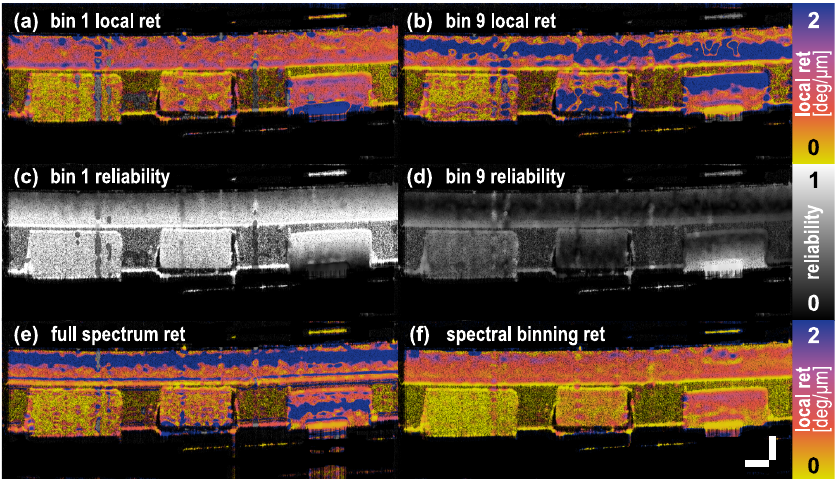}
\caption{\label{fig:fig5}  Inaccurate estimation of local retardation can be avoided with a small amount of polarization mode dispersion (PMD) in combination with spectral binning. (a)-(f) Cross-sectional images of a two-layer phantom. (a) Local retardation reconstructed using only the 1st spectral bin. (b) Local retardation reconstructed using only the 9th spectral bin. (c,d) Reliability metric maps of the 1st and 9th spectral bin, respectively. (e) Local retardation reconstructed using the entire spectrum without spectral binning. (i) Local retardation image reconstructed with spectral binning combining all bins. Scale bars in (f) measure \SI{100}{\mu m} (vertical) and \SI{400}{\mu m} (horizontal).}
\end{figure}

Coupling the sample light through the 1.5-\si{\meter}-long single mode fiber twisted around the polarization controller paddles provided sufficient PMD for our broad-bandwidth source. For spectral binning, we multiplied the spectral fringe signals with Hanning windows $h(k,n)$ of width $\Delta k/N$ centred on $n\cdot\Delta k/\left(2N\right)$ within the available $k$-support, $\Delta k$, $n\in\left[ 1,2N-1\right]$, $N=5$, resulting in 9 spectral bins, to compute the binned Stokes vectors ${\mathbf{r}}\left(z,n\right)$. We also evaluated the degree of polarization ${\text{DOP}}=\left\langle {{{{\left( {{Q^2} + {U^2} + {V^2}} \right)}^{1/2}}/I}} \right\rangle$, where $\left\langle \right\rangle$ indicates averaging over the spectral bins, and $Q$, $U$, $V$ and $I$ are the spatially filtered Stokes components before normalization. Following the identical processing for local retardation for each bin as described above, we obtained the rotation vectors ${\boldsymbol{\upbeta}}\left( z,n\right)$. Fig.~\ref{fig:fig5}(a-d) illustrates the local retardation of bins 1 and 9, together with a map $w(z,l) = \left|  {\mathbf{t}} - {\mathbf{m}} \right|$ expressing the reliability of the given Stokes vector by the distance from its mirror state, for a tissue-like birefringence phantom. Bin 9 results in high local retardation values but with little reliability, unlike bin 1, which indicates more modest local retardation yet with higher reliability. The ${\boldsymbol{\upbeta}}\left( z,n\right)$ with high reliability of all bins describe the same sample retardation but may be offset in their relative orientation due to system PMD. The required rotation ${\mathbf{G}}(n)$ to align the vectors of each bin to the central bin $N$ in the least-square sense is given by:

\begin{equation}
\max _ { {\mathbf{G}}(n) } \operatorname { Tr } \left( {\mathbf{G}}(n) \cdot \sum _{z,l} {\boldsymbol{\upbeta}} (z,l,N) \cdot {\boldsymbol{\upbeta}}^{\top} (z,l,N) \cdot w(z,l) \right)
\label{eqn:eq5}
\end{equation}

\noindent where $z$ and $l$ are point indices in the axial and lateral directions, respectively, ${\mathbf{G}}(n)$ is assumed constant within an entire B-scan, and the sum is taken over all points with sufficient ${\text{DOP}}>0.8$ and signal intensity ${\text{SNR}}>\SI{5}{dB}$. From the singular value decomposition of the $3\times3$ matrix defined by the summation $\sum {\boldsymbol{\upbeta}}\cdot{\boldsymbol{\upbeta}}^\top\cdot w = {\mathbf{U}}\cdot{\mathbf{W}}\cdot{\mathbf{V}}^\dagger$, where $^\dagger$ denotes conjugate transpose, the solution to Eq.~(\ref{eqn:eq5}) is obtained by ${\mathbf{G}}={\mathbf{V}}\cdot{\mathbf{U}}^\dagger$. Lastly, the aligned rotation vectors are averaged among the spectral bins considering their weights $w(z,l)$, and then axially filtered, as previously, to obtain the final local retardation image, free from artifacts, as demonstrated in Fig.~\ref{fig:fig5}(f).

In conclusion, we demonstrated the peculiar properties of the mirror polarization state that manifest when measuring backscattered light along identical illumination and detection paths free of polarization-dependent loss. In PS-OCT, the mirror state constrains the evolution of the depth-dependent polarization states and enables local retardation imaging, which previously has not been available to PS-OCT without substantially more complex measurements using multiple input states.

\begin{acknowledgments}
Support is acknowledged from a National Research Foundation Singapore (NRF-CRP13-2014-05), Ministry of Education Singapore (MOE2013-T2-2-107 \& RG 83/18 (2018-T1-001-144)), and NTU-AIT-MUV program in advanced biomedical imaging (NAM/15005), and part by the National Institutes of Health grants P41EB-015903, R03EB-024803.
\end{acknowledgments}

\nocite{*}


\providecommand{\noopsort}[1]{}\providecommand{\singleletter}[1]{#1}%

\end{document}